# Moiré superlattices in strained graphene-gold hybrid nanostructures

*András Pálinkás*[1,3], *Péter Süle*[1], *Márton Szendrő*[1], *György Molnár*[1], *Chanyong Hwang*[2,3], *László P. Biró*[1,3]*, and Zoltán Osváth*[1,3,*]

[1]Institute of Technical Physics and Materials Science, MFA, Centre for Energy Research, HAS, 1525 Budapest, P.O. Box 49, Hungary

[2]Center for Nano-metrology, Division of Industrial Metrology, Korea Research Institute of Standards and Science, Yuseong, Daejeon 305-340, Republic of Korea

[3]Korea-Hungary Joint Laboratory for Nanosciences (KHJLN), P.O. Box 49, 1525 Budapest, Hungary

**Abstract**

Graphene-metal nanoparticle hybrid materials potentially display not only the unique properties of metal nanoparticles and those of graphene, but also additional novel properties due to the interaction between graphene and nanoparticles. This study shows that gold nanoislands can be used to tailor the local electronic properties of graphene. Graphene on crystalline gold nanoislands exhibits moiré superlattices, which generate secondary Dirac points in the local density of states. Conversely, the graphene covered gold regions undergo a polycrystalline → Au(111) phase transition upon annealing. Moreover, the nanoscale coexistence of moiré superlattices with different moiré periodicities has also been revealed. Several of these moiré periodicities are anomalously large, which cannot be explained by the standard lattice mismatch between the graphene and the topmost Au(111) layers. Density functional theory and molecular dynamics simulations show for the first time that in such cases the graphene and the interfacial metallic layer is strained, leading to distorted lattice constants, and consequently to reduced misfit. Room temperature charge localization induced by a large wavelength moiré pattern is also observed by scanning tunneling spectroscopy. These findings can open a route towards the strain engineering of graphene/metal interfaces with various moiré superlattices and tailored electronic properties for nanoscale information coding.

---

[*]Corresponding author. Tel: +36 1 392 2222 / 3616. E-mail: zoltan.osvath@energia.mta.hu (Zoltán Osváth).

1. **Introduction**

Tailoring the atomic and electronic structure of graphene [1–3] e.g. via strain engineering [2] has been a subject of intense research, due to its possible applications in nanoelectronics. Ultra-narrow graphene nanoribbons fabricated by scanning tunnelling lithography have electronic bandgaps up to 0.5 eV [4] and can show electron-electron interaction-induced spin ordering along their edges, depending on the nanoribbon edge structure [5]. Recent theoretical investigations [6] and scanning tunnelling microscopy (STM) experiments on supported graphene [7] revealed that the local density of states (LDOS) can be also modified by periodic potentials which create secondary Dirac points (SDP). Such periodic potentials can be induced in graphene by moiré patterns using crystalline substrates like h-BN [8,9]. The lattice mismatch and relative rotation between graphene and the crystalline substrate determine the period of the moiré pattern, and consequently the energy at which the SDPs appear. A similar effect is observed also for twisted bilayer graphene, where van Hove singularities appear as a function of the rotation angle between the two layers [10,11]. Moiré superlattices develop on metal supported graphene as well [12]. Depending on how strongly the graphene layer interacts with the metal substrate, its band structure can be strongly modified (e.g. Ru, Ni) [13,14] or it can keep the linear dispersion characteristic (e.g. Ir, Pt) [15,16]. STM and tunnelling spectroscopy (STS) studies on graphene on Ru(0001) – corresponding to strong interaction – showed spatial modulation in the LDOS [17,18] and electron localization [18] imposed by the formation of a moiré pattern. On the other hand, it has been shown by angle-resolved photoelectron spectroscopy of the weakly interacting graphene on Ir(111) [15] system, that the periodic potential associated with the moiré pattern can give rise to Dirac cone replicas and the opening of minigaps in the graphene band structure. Recently it has been shown that graphene grown [19] or transferred [20] onto Au(111) also forms moiré structures. Convex and concave graphene moiré superlattices can coexist on Au(111), although the concave (depressions) morphology is dominant [21]. Gold is commonly used to electrically contact graphene for device fabrication. These contacts are made by sputtering or e-beam evaporation and consist of polycrystalline grains. The interaction of graphene with such a polycrystalline gold layer was investigated recently by transferring graphene flakes on a sputtered, 8 nm gold layer [22]. Based on STS measurements, it was concluded that the gold substrate is the mixture of (111) crystallites and amorphous-type regions. The graphene-gold interaction, and generally the metal-graphene interaction in previous works, was defined by the growth or transfer process. Here we show that the strength of graphene-gold interaction can be tuned by annealing graphene-covered gold nanoislands at moderate temperatures (well below the graphene growth temperatures) and that graphene plays a key role in the formation of crystalline gold surfaces, i.e. in the recrystallization of amorphous-type Au. The observed large wavelength moiré patterns involve (111) gold surfaces with modified lattice parameter, which enable spatial modulations and the emergence of secondary Dirac points in the LDOS of gold-supported graphene. We demonstrate that in such strained moiré superlattices the convex (protrusion) morphology dominates.

## 2. Methods

*2.1 Preparation of graphene-gold hybrid nanostructures*

Graphene-gold hybrid nanostructures were prepared as follows: gold grains of 99.99% purity were applied as source material for evaporation, which was carried out from an electrically heated tungsten boat, at a background pressure of $5\times10^{-7}$ mbar. Thin gold films of 5 nm and 8 nm were evaporated onto a graphite (HOPG) substrate at a rate of 0.1 nm/s. After evaporation the samples were annealed in argon atmosphere at 400 $^{o}$C for 1 hour, which resulted in the formation of gold nanoislands with heights of 15-20 nm and lateral dimensions of several hundreds of nanometers.

Large-area graphene was grown on a mechanically and electro-polished copper foil (25 μm thick, 99.8% purity, Alfa-Aesar), which was inserted into a thermal chemical vapour deposition furnace. The furnace was evacuated to ~$10^{-4}$ mbar and the temperature was raised to 1010 $^{o}$C with $H_2$ gas flow (~$10^{-2}$ mbar). When the temperature became stable, both $CH_4$ (20 sccm) and $H_2$ (5 sccm) were injected into the furnace for 8 minutes to synthesize the graphene. After the growth, we cooled down the furnace with a cooling rate of 50 $^{o}$C/min.

The graphene sample was transferred onto the gold nanoislands using thermal release tape, and an etchant mixture consisting of $CuCl_2$ aqueous solution (20%) and hydrochloric acid (37%) in 4:1 volume ratio. After the etching procedure, the tape holding the graphene was rinsed in distilled water, then dried and pressed onto the HOPG surface covered by the gold. The tape/graphene/gold/HOPG sample stack was placed on a hot plate and heated to 95 °C, slightly above the release temperature of the tape. The tape was removed, leaving behind the graphene on top of the gold nanoislands (and on the HOPG). A second annealing was performed at 400 °C in argon atmosphere for 30 minutes, to improve the adhesion of graphene to the nanoislands. The graphene-covered gold nanoislands were investigated by STM and STS, using a DI Nanoscope E operating under ambient conditions. The STM measurements were performed in constant current mode. Finally, a third annealing was also performed at 650 °C for two hours.

*2.2 Density functional theory (DFT) and classical molecular dynamics (CMD) simulations*

In order to reproduce the experimentally found moiré phases, density functional theory (DFT)-adaptive classical molecular dynamics (CMD) simulations were performed. CMD has been used as implemented in the LAMMPS code (Large-scale Atomic/Molecular Massively Parallel Simulator) [23]. Periodic triclinic (rhombic) simulations cells have been constructed with different sizes ranging from 100 x 100 up to 340 x 340 graphene honeycomb unit cells. In the largest case the simulation cell includes few millions of atoms. 20-30 layers of Au(111) were used underneath the graphene overlayer together with 3 fixed bottom layers which are necessary to avoid the waving of the support as well as the rotation and the translation of the system. The systems are carefully matched at the cell borders in order to give rise to perfect periodic systems. Non-periodic structures have also been optimized by simple minimizers which also provide moiré patterns. Further refinement of the periodic pattern can be obtained by time-lapsed CMD. Isothermal constant volume (NVT) and isobaric-isothermal (NPT ensemble) simulations (with Nose-Hoover thermostat and a pressostat) were carried out at 300 K. Vacuum regions were inserted above and below the slab of the graphene-substrate system to ensure the periodic conditions not only in lateral directions ($x,y$) but also in the direction perpendicular to the graphene sheet ($z$), which is required by the NPT scheme as implemented in LAMMPS. The variable time step algorithm has been exploited. The OVITO code has been used to display the obtained structures and to plot the colour-coded topographic images [24]. The AIREBO (Adaptive Intermolecular Reactive Bond Order) potential has been used for the graphene sheet [25]. For Au(111), the modified embedded-atom method (MEAM) is employed [26]. For the cross-interactions at the interface (C-Au) a new DFT-adaptive bond order potential has been developed in a similar way as it has been done for C-Cu [27,28]. Essentially an Abell-Tersoff angular dependent force field [29,30] has been fitted to accurate van der Waals (*vdw*) DFT calculations [31]. The fully periodic SIESTA code [32] has been used for DFT calculations together with the DF2 nonlocal *vdw* functional. DFT calculations have also been used for geometry optimization of small model graphene/Au(111) systems in order to demonstrate that distorted lattices survive DFT geometry relaxation processes.

3. **Results and Discussion**

Graphene-covered gold nanoislands were fabricated on highly oriented pyrolytic graphite (HOPG) substrates, as described above. The obtained gold nanoislands are irregularly shaped, with heights of 15 – 20 nm. Their lateral dimensions vary from 100 nm to about 400 nm in width, and from 150 nm to 1.2 µm in length (Fig. 1a). Graphene wrinkles develop and connect neighbouring nanoislands (Fig. 1a, white arrows), as observed by atomic force microscopy (AFM).

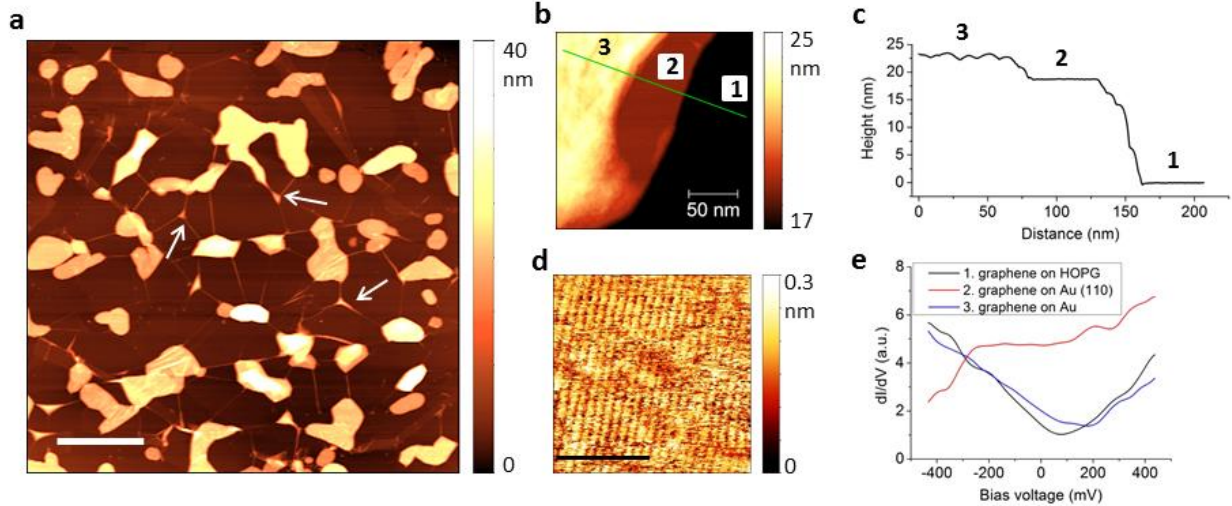

***Fig. 1.*** *Fabrication of graphene-covered gold nanostructures. (**a**) AFM image of graphene-covered gold nanoislands prepared on HOPG substrate. Graphene wrinkles are marked with white arrows. Scale bar is 1 µm. (**b**) STM image of a graphene-covered gold nanoparticle. Tunneling parameters: U = 0.5 V, I = 1 nA. Three different regions can be distinguished: 1 – graphene on HOPG, 2 – graphene on crystalline gold, and 3 – graphene on non-recrystallized gold grains. The line section corresponding to the green solid line is shown in (**c**). (**d**) Unreconstructed Au(110) surface measured in region 2 (U = 0.1 V, I = 1 nA). Scale bar is 4 nm. (**e**) dI/dV spectra measured in the regions marked 1 – 3 in (**b**) and (**c**). Graphene loses its electronic properties and shows metallic characteristics (red line) when lies on Au(110). The less ordered, polycrystalline gold only increases the p-doping of graphene (blue line), compared to graphene on HOPG (black line).*

STM measurements performed on graphene-covered individual nanoislands after annealing at 400 °C reveal two distinct regions (Fig. 1b): a smooth region (marked *2*), with surface root mean square roughness comparable to that of HOPG (RMS = 0.06 nm), and a rougher region (marked *3*) with RMS of 0.35 nm (see also the line section in Fig. 1c), typical for as-deposited metallic films consisting of small grains. A higher resolution STM image acquired in the smooth region is shown in Fig. 1d, where we observe parallel lines with period of 0.41 nm. This corresponds to the distance between the atomic rows of an unreconstructed Au(110) surface. *dI/dV* spectra measured in this region (Fig. 1e, red line) show metallic density of states, and the onset of a Shockley surface state appears at -350 meV. The observed surface state is upshifted compared to the clean, unreconstructed Au(110) [33], similarly to the case when a monolayer of Ag atoms is adsorbed onto the Au(110) [33]. Note that the spectra are measured on graphene-covered gold, and the characteristics of crystalline gold are observed through the graphene. This indicates strong graphene-gold interaction in this region, where graphene characteristics are lost due to the hybridization of carbon π-states with the metallic substrate [34]. In contrast, when on top of non-recrystallized gold grains (region *3*), graphene is sufficiently decoupled and the linear LDOS is preserved (Fig. 1e, blue curve), reflecting a weak graphene-gold interaction. Here, the electrostatic *p*-doping from gold is the dominant effect. The Dirac point of graphene is shifted with 130 meV compared to the Dirac point of HOPG-supported graphene (Fig. 1e, black curve), measured on the region marked *1* in Fig. 1b-c. We note that reconstructed Au(111) surfaces were also observed on graphene-covered gold nanoislands annealed at 400 $^o$C (not shown). However, these crystalline surfaces were less than 20% of the total graphene-covered gold areas.

The formation of crystalline gold domains under graphene is an intriguing phenomenon which should be related to the reconstruction of surface gold atoms during annealing. To investigate this hypothesis, we further annealed the graphene-covered gold nanoislands at higher temperatures (650 $^o$C), in order to increase the size of crystalline regions. Surprisingly, subsequent STM investigations revealed that all (100%) of the observed graphene-covered gold surfaces became (111) crystalline domains during annealing at 650 $^o$C. This is illustrated in Fig.

2, with STM images measured on a graphene-covered gold nanoisland. More details about the shape and dimensions of the nanoislands formed during annealing at 650 °C are shown in the supplementary data (Fig. S1).

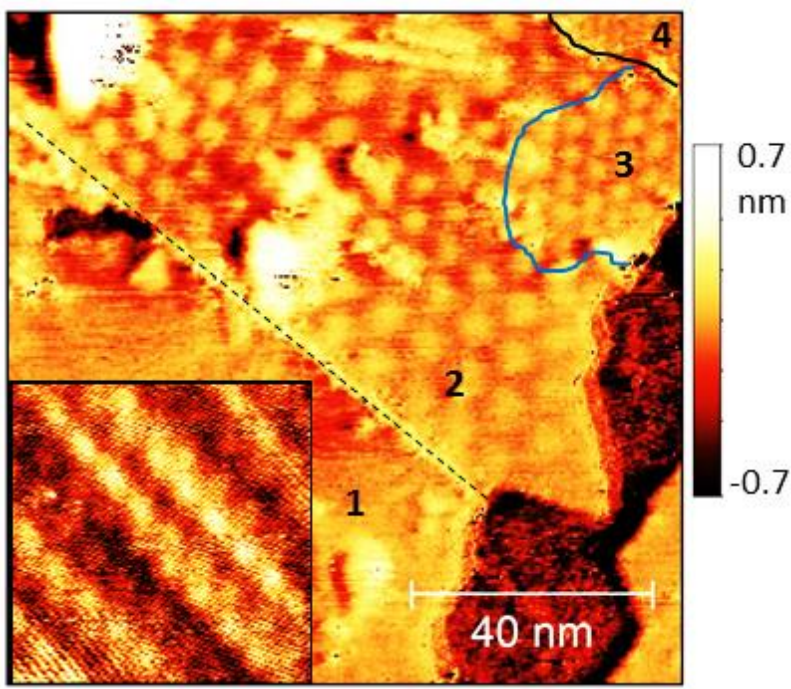

*Fig. 2. STM image of a hybrid graphene-gold nanocrystal after annealing at 650 °C. All graphene-covered gold surfaces are crystalline after annealing (U = 0.5 V, I = 1 nA). The areas denoted 1 – 4 are separated by graphene grain boundaries (dashed line, blue line, and black solid line, respectively). Moiré superlattices of different wavelengths are observed. Area 1: moiré of 1.9 nm superimposed on Au(111) surface reconstruction (see inset). Area 2: moiré of 7.7 nm. Area 3: moiré of 5.1 nm. Area 4: moiré of 2.5 nm (see Fig. 4).*

The light-coloured areas denoted *1 – 4* in Fig. 2 correspond to gold-supported graphene, while areas with darker contrast correspond to bare gold surfaces. Several graphene grain boundaries are observed at the junction between areas *1-2*, *2-3*, and *3-4*, which are emphasized with dashed line, blue line, and black solid line, respectively. Moiré patterns are formed on all numbered areas, while in each area the moiré pattern has different wavelength. In area *1*, the "herringbone" surface reconstruction of Au(111) is observed (period of ~6.3 nm) [35] (see the inset in Fig. 2), which has its origin in a uniaxial contraction of ~4.4% of the surface layer with respect to the bulk. A high resolution STM image measured in area *1* is shown in Fig. 3a.

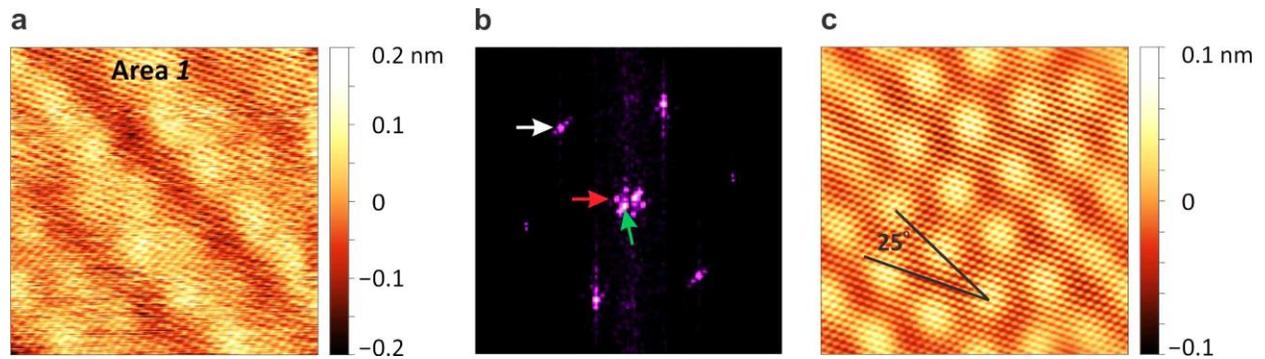

*Fig. 3. High resolution STM image (**a**) of area labelled 1 in Fig. 2 (U = 0.1 V, I = 1 nA). (**b**) 2D Fourier-transform image of (a). Bright maxima are related to the atomic periodicity (the six most distant spots from image centre, one is highlighted by white arrow), moiré pattern (the six spots closer to the image centre, one is highlighted by red arrow), and the herringbone reconstruction (two brighter spots closest to the image centre, one highlighted by green arrow). (**c**) Fourier-filtered image of (a) showing the moiré superlattice of 1.9 nm and the atomic lattice of graphene. A moiré angle of 25$^o$ is measured.*

The two-dimensional Fourier-transform of this image (Fig. 3b) reveals the atomic periodicity (white arrow), the moiré period (red arrow), and the herringbone reconstruction as well (green arrow). Fig. 3c is a Fourier-filtered image where the atomic lattice of graphene and a moiré pattern with periodicity of 1.9 nm is clearly visible. Here, a moiré angle of $25° \pm 1°$ is measured between the zig-zag orientation of graphene and the line determined by the moiré humps. In area *2* and *3*, large wavelength moiré patterns of 7.7 nm and 5.1 nm are formed, respectively, while in area *4* a moiré of 2.5 nm is observed. The corresponding high resolution STM images performed in these areas are shown in Fig. 4.

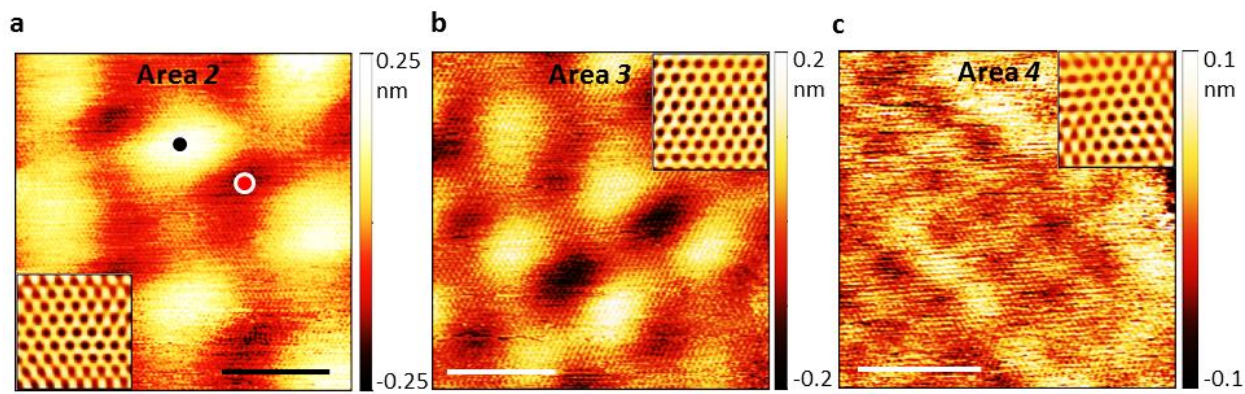

*Fig. 4. High resolution STM images measured in the areas denoted 2 – 4 in Fig. 2, showing moiré superlattices of different wavelengths (U = 0.1 V, I = 1 nA). (a) Area 2: moiré of 7.7 nm. dI/dV spectra measured at the positions marked by dots are presented in Fig. 6. (b) Area 3: moiré of 5.1 nm. (c) Area 4: moiré of 2.5 nm. Scale bars are 5 nm. The insets show the honeycomb graphene lattice.*

Although there is no herringbone reconstruction in areas *2-4*, the hexagonal symmetry of the moiré superlattices imply that Au(111) surfaces developed on these areas as well. Note, that the gold surfaces not covered with graphene do not display single-crystalline surface, but rather a disordered one (Fig. S2b). This shows very clearly that the recrystallization of gold surfaces is induced by the graphene overlayer.

The various experimentally found moiré patterns were successfully reproduced by classical molecular dynamics (CMD) simulations, as shown with the colour coded topographic images in Fig. 5.

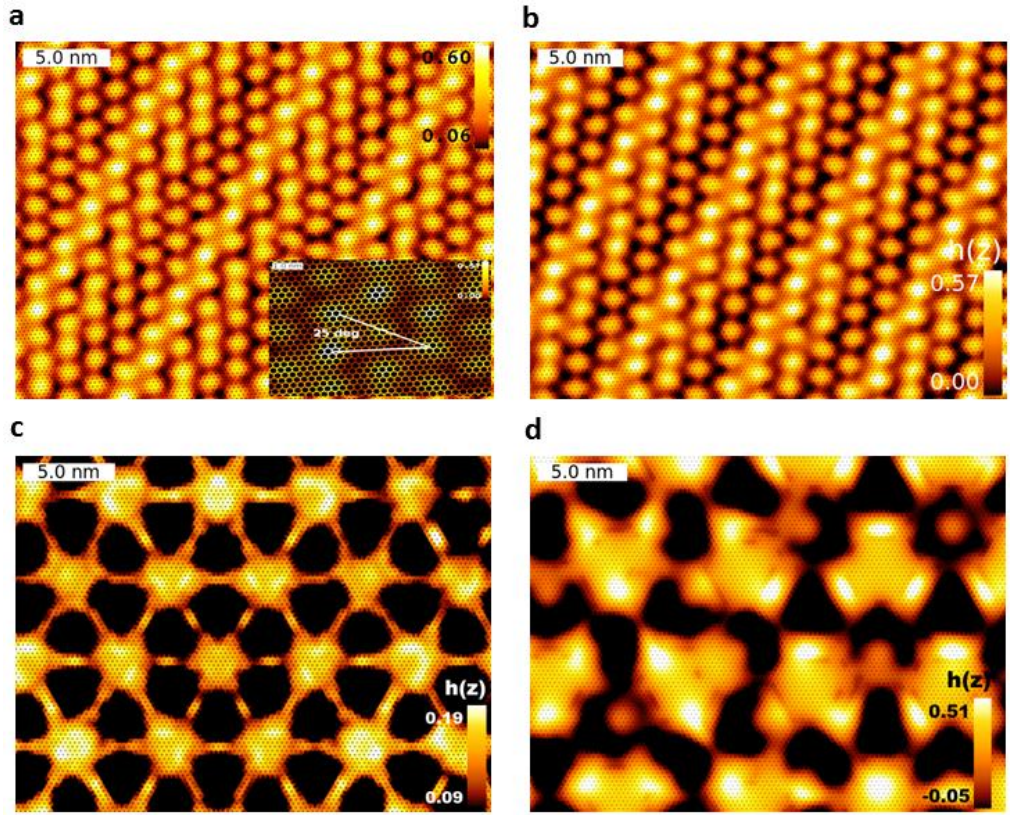

*Fig. 5. Topographic images of moiré superlattices with different periodicity as obtained by DFT-adaptive classical molecular dynamics simulations. The misfit in various gr/Au(111) systems has been tuned systematically in order to approach the experimentally obtained moiré periodicity. (a) moiré of 1.8 nm on Au(111) ($d_{CC}$ = 2.46 Å, $d_{AuAu}$ = 2.82 Å). The simulated moiré angle of 25° is shown in the inset. (b) moiré of 2.4 nm superlattice ($d_{CC}$ = 2.49 Å, $d_{AuAu}$ = 2.74 Å). (c) moiré of 5.2 nm ($d_{CC}$ = 2.55 Å, $d_{AuAu}$ = 2.68 Å). (d) moiré of 7.8 nm ($d_{CC}$ = 2.5 Å, $d_{AuAu}$ = 2.61 Å). The lattice constants of the graphene layer and the topmost Au(111) layer are given in parenthesis.*

A new DFT-adaptive interface interaction potential has been developed which is able to describe adequately the weak van der Waals forces at the graphene/Au(111) interface, as it has already been demonstrated for the graphene/Cu(111) system [27,28]. The overall performance of this force field is superior to pairwise potentials such as the simple Lennard-Jones one [27] at the interface. According to the rigid lattice approximation [36], the moiré periodicity is inversely proportional to the lattice misfit between the overlayer graphene and the support. Several experimental moiré periodicities and the corresponding lattice misfits are summarized in Fig. S6, for various graphene/support systems: graphene/hBN [9], graphene/Cu(111) [37,38], graphene/Ru(0001) [39], graphene/Rh(111) [40], graphene/Ir(111) [41], graphene/Pt(111) [16], graphene/SiC [42]. Using the equilibrium interatomic distances of bulk Au (2.88 Å) and graphene (2.46 Å), a maximum moiré period of only 1.8 nm can be formed in graphene/Au(111) (at zero rotation angle), due to the significant misfit. This implies that large moiré periodicities (5.1 nm, 7.7 nm) can only be explained by considerable lattice distortions both in the graphene and in the support layers. The larger the moiré periodicity is the smaller lattice misfit is necessary. These in principle are serious distortions, especially in the support. However, we find that the graphene/Au(111) system is peculiar, because it can adjust itself relatively easily in order to build up such an unexpected superlattice arrangement with non-equilibrium interatomic distances. Our CMD simulations support the rigid lattice approximation [36] and provide moiré phases (Fig. 5) with periodicities similar to that found in STM experiments (Fig. 3 and Fig. 4).

In particular, we are able to reproduce the moiré periodicity and the convex (protruding) topography. We showed recently, that both the convex and concave (nanomesh) graphene moiré morphologies are stable [21], and that graphene on Au(111) generally favours the formation of nanomesh-like morphology [21]. The unexpected convex moiré patterns observed in this work are related to strain appearing in both graphene and the top gold atomic lattices, as discussed below. In Fig. 5a we are able to reproduce the experimentally observed periodicity of 1.8 nm and moiré angle of 25° (see inset) using standard lattice constants. Furthermore, the anomalously large wavelength moiré superlattices are also reproduced, as shown in Fig. 5b-d. However, CMD simulations reveal that in these cases the standard lattice constants do not apply, the crystal lattices are considerably distorted. In particular, we find that if the graphene layer expands by 1-3 % (from 2.46 Å up to 2.55 Å) while the topmost layer of Au(111) shrinks by 6-7 % (from the herringbone 2.82 Å down to 2.61 Å). The misfit is reduced from 12.8 % to 4.2 % and the moiré periodicity increases from 1.8 nm up to 7.8 nm. The obtained lattice constants and other moiré characteristics are summarized in Table 1.

|  | $\lambda$ (nm) | moiré angle (°) | $d_{CC}$ (Å) | $d_{AuAu}$ (Å) | lattice misfit (%) | rotation angle (°) |
| --- | --- | --- | --- | --- | --- | --- |
| **Area 1** (Fig. 2) | 1.9 ± 0.1 | 25 ± 1 | 2.46 | 2.82 | 12.8 | 3.9 |
| **Area 4** (Fig. 2) | 2.5 ± 0.1 | 28 ± 1 | 2.49 | 2.74 | 9.1 | 3.1 |
| **Area 5** (Fig. S2a) | 3 ± 0.1 | 24 ± 1 | 2.47 | 2.65 | 6.8 | 0.9 |
| **Area 3** (Fig. 2) | 5.1 ± 0.1 | 28 ± 1 | 2.55 | 2.68 | 4.8 | 1.5 |
| **Area 2** (Fig. 2) | 7.7 ± 0.1 | 28 ± 1 | 2.5 | 2.61 | 4.2 | 1.3 |

***Table 1.*** *Moiré characteristics: experimental moiré wavelength (λ) and moiré angles, $d_{CC}$ and $d_{AuAu}$ – graphene and gold lattice constants, respectively, obtained from CMD simulations. The rotation angles are calculated according to Eq. S1 (see supplementary data).*

The energy cost of this non-equilibrium process is surprisingly small and amounts to 0.1±0.03 eV/atom, which can be provided by thermal motion. The considerable contraction of the topmost layer of Au(111) requires as small as < 0.05 eV/atom energy investment. Therefore we find that the graphene/Au(111) system builds up stable non-equilibrium moiré superlattices during annealing, with considerably distorted lattice constants. The mechanism that we propose for the formation of anomalously large wavelength moiré phases is related to the cooling process. Gold atoms are mobile at 650°C and reorganize on the surface. The hexagonal structure of graphene acts as a template and facilitates the formation of Au (111) surfaces. As the temperature decreases, the top gold layer compresses and the herringbone surface reconstruction appears. Further compression of the gold occurs during the cooling process, while graphene expands due to its negative thermal coefficient [43]. Anomalously large moiré patterns form as the misfit between graphene and gold lattice constants becomes sufficiently small. Additionally, covalent bonds can develop at graphene edges and grain boundaries between carbon and gold atoms, which stabilize the anomalous moiré phases by hindering further rotation or translation.

DFT geometry optimization supports the stretching of the graphene layer (Fig. S4 and Fig. S5). Moreover, DFT calculations also show that the Au-Au interatomic distances in the topmost Au(111) layer contract to ~2.65 Å, in agreement with CMD simulations. Additionally, we show that the Au(111) topmost layer reconstructs according to the moiré pattern (see Fig. S4b and Fig. S5b). This agrees with our previous findings on nanomesh-type moiré superlattices [21]. It is important to note that the topography of the superlattice is extremely sensitive to strain. In the case of a slightly stressed system (the rms forces of the entire system is ~0.001 eV/Å) the convex morphology occurs. In the fully relaxed system (rms forces < 0.0001 eV/Å) the purely convex pattern turns into a mixed morphology, which displays both convex and concave features (Fig. S3b). We find that the energy difference between the two morphologies is less than 0.005 eV/atom, and the mixed morphology is slightly more favourable. Therefore, tiny distortions in the lattice can transform the most stable concave superlattice into a convex one. In the case of larger moiré wavelengths of 2.5 and 3.1 nm (Fig. S4 and Fig. S5) the convex morphology seems to be the unique morphology after geometry relaxation procedure. In these cases the crystal lattices are distorted: graphene is stretching up to 2.49-2.56 Å and the topmost Au(111) layer becomes contracted with lattice constants of 2.65-2.8 Å.

Tunnelling spectra were acquired on the moiré patterned graphene/Au(111) areas. The corresponding dI/dV curves display secondary Dirac points (SDP) in the LDOS. The energy of SDPs depends on the moiré wavelength, as shown in Fig. 6a-b.

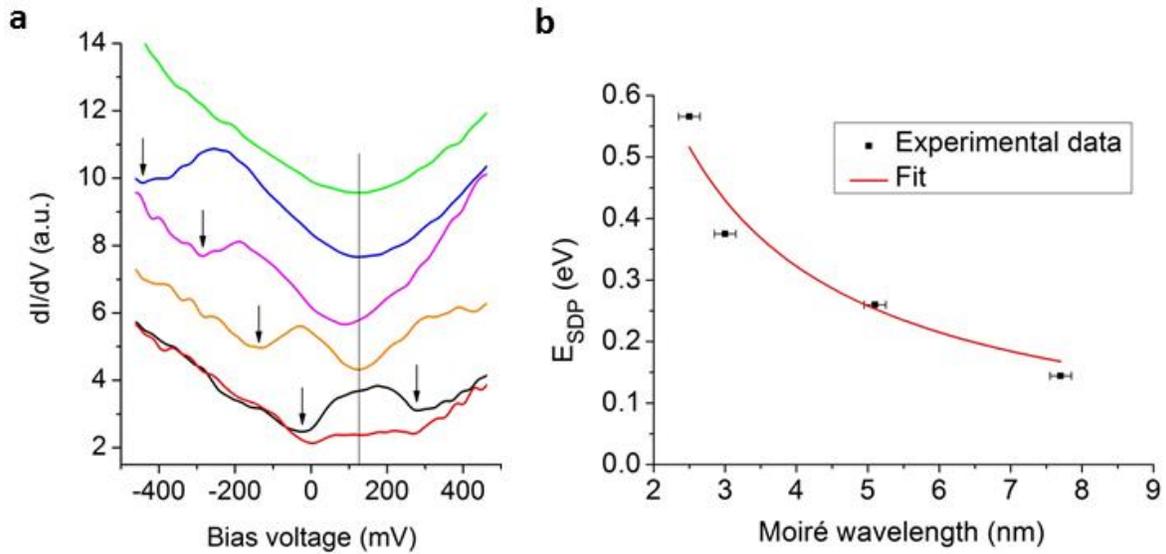

*Fig. 6. Local density of states of graphene on Au (111) showing superlattice Dirac points. (**a**) Experimental dI/dV curves for five different moiré wavelengths: 1.9 nm (green, area 1 in Fig. 2), 2.5 nm (blue, area 4 in Fig. 2), 3.0 nm (magenta, Fig. S2a), 5.1 nm (orange, area 3 in Fig. 2), and 7.7 nm (area 2 in Fig. 2). For this latter, spectra measured on both topographically high (black) and low (red) positions are shown. These positions are marked in Fig. 2b with black and red dots, respectively. The vertical line marks the approximate position of the main Dirac point. The secondary dips in the spectra are marked by arrows. Each spectrum is an average of 8 measurements. (**b**) Energy of the secondary dips measured from the Dirac point, as a function of*

*moiré wavelength. The black symbols are experimentally measured values, while the red line is the theoretical fit to the data.*

The experimentally measured values are fitted with the expected theoretical dependence [8] $E_{SDP} = 2\pi\hbar v_F/(\sqrt{3}\lambda_M)$, where $\lambda_M$ is the moiré wavelength. Best fit is obtained for the Fermi velocity $v_F = 0.54 \times 10^6\ ms^{-1}$. Note, that for the moiré wavelength of 1.9 nm (Fig. 6a, green line, area *1* in Fig. 2) the expected SDP (0.68 eV) falls off the bias voltage range used for spectroscopy, and thus is not observed.

As already mentioned above, lattice distortions are necessary to produce large moiré wavelengths, and hence SDPs in the ±0.5 eV range from the Dirac point. It should be emphasized that if any finite rotation angle between graphene and Au(111) is to be considered, then even more significant lattice parameter changes are needed in order to reproduce the experimentally found superlattices, since the moiré period decreases with increasing rotation angle [28,36]. Moreover, further lattice distortions are energetically unfavourable.

Finally, for the moiré of 7.7 nm, not only SDPs, but room-temperature charge localization is also observed on the topographically high positions (Fig. 6a, black line) of the moiré pattern. Until now, similar localization near the Dirac point was only observed at low temperatures for twisted graphene layers [10,11]. The CMD simulations of this large wavelength moiré (Fig. 5d) result in a maximum geometric corrugation of $h = h_{max} - h_{min} \cong 0.5$ Å, where $h_{max}$ and $h_{min}$ are the highest and lowest carbon atom positions, respectively. On the other hand, the maximum corrugation of the 7.7 nm moiré measured by STM (Fig. 4a) is $z = z_{max} - z_{min} \cong 2$ Å, with $z_{max}$ and $z_{min}$ the height values of topographically high and low positions, respectively. The STM image was measured at U = 100 mV, which is in the bias range where the charge localization was observed. Here, the LDOS of topographically high positions is significantly higher than the LDOS of topographically low positions (Fig. 6a, black and red spectra, respectively), which implies additional upwards *z*-movement of the STM tip in order to keep the current constant. This gives an electronic contribution to the measured corrugation of about 1.5 Å ($z - h$). Thus, the observed charge localization supports the convex (protrusion) character of the moiré superlattice. Furthermore, based on the geometric corrugation determined by CMD simulations, the strain in this moiré structure is $h/l = 1.3\%$, with $l = 39$ Å the half moiré wavelength. This is a relatively low strain and we think that the effect of possible strain-induced pseudo-magnetic fields [44, 45] on the LDOS is negligible.

## 4. Conclusions

In summary, we demonstrated that the superlattice moiré periodicity – and the graphene LDOS – can be tuned not only by rotation (misorientation), as shown previously [8, 28], but also by tuning the lattice mismatch between graphene and the topmost layer of the support. Annealing of graphene/gold nanostructures gives rise to the formation of genuine graphene-gold hybrid nanocrystals where graphene is stretched and the interface gold layer is considerably contracted. We revealed that graphene induces the recrystallization of the polycrystalline Au-surface into reconstructed Au(111) surface. Moreover, the graphene moiré pattern induces additional reconstruction of the Au(111) surface, as demonstrated by large scale DFT calculations. Using DFT-adaptive CMD simulations we developed a simple model with which the observed anomalously large moiré periodicities could be explained. We revealed that the moiré periodicity can be tuned by the formation of discrete graphene/top gold layer lattice parameter pairs. The topography of the moiré superlattices shows some additional peculiarities: instead of the expected concave curvature [21] (depressions) convex (protrusion) morphology has been identified, which is attributed to the built-in stress in the superlattice. Even in the fully relaxed graphene/Au(111) system with equilibrium superlattice we find that the morphology is extremely sensitive to small distortions in the crystal lattices. Our findings open up new avenues for the nanoscale modulation of the electronic properties of graphene by strain engineering, and for the controlled recrystallization of Au nanostructures.

**Acknowledgments**

The research leading to these results has received funding from the People Programme (Marie Curie Actions) of the European Union's Seventh Framework Programme under REA grant agreement n° 334377, and from the Korea-Hungary Joint Laboratory for Nanosciences. The OTKA grants K-101599 and K-112811, as well as the NKFIH project TÉT_12_SK-1-2013-0018 in Hungary are acknowledged. Z. Osváth acknowledges the János Bolyai Research Fellowship from the Hungarian Academy of Sciences. The DFT calculations and CMD simulations were done on the supercomputers of the NIIF Supercomputing Center in Hungary. C. Hwang acknowledges funding from the Nano-Material Technology Development Program (2012M3A7B4049888) through the National Research Foundation of Korea (NRF) funded by the Ministry of Science, ICT and Future Planning.

**Appendix A. Supplementary data**

Supplementary data related to this article can be found at http://dx.doi.org/10.1016/j.carbon.2016.06.081.